\documentclass[pre,10pt]{revtex4}

\usepackage{amsmath}    
\usepackage{graphicx}   
\usepackage{verbatim}   
\usepackage{color}      
\usepackage{subfigure}  
\usepackage{hyperref}   

\begin{document}

\title{The influence of local field corrections 
       on Thomson scattering in non-ideal two-component plasmas}

\author{Carsten Fortmann, August Wierling, and Gerd R\"opke}
\email{august.wierling@uni-rostock.de}
\affiliation{Universit\"at Rostock, Institut f\"ur Physik,
             18051 Rostock, Germany}
\date{\today}

\begin{abstract}

Thomson scattering in non-ideal (collision-dominated)
two-component plasmas is calculated accounting for electron-ion
collisions as well as electron-electron correlations. This is achieved
by using a novel interpolation scheme for the electron-electron
response function generalizing the traditional Mermin approach. Also,
ions are treated as randomly distributed inert scattering centers. 
The collision frequency is taken as a dynamic and complex quantity
and is calculated from a microscopic quantum-statistical approach.
Implications due to different approximations for the electron-electron
correlation, i.e. different forms of the OCP local field correction,
are discussed.

\end{abstract}

\maketitle

\section{Introduction}

Recently, Thomson scattering has been established as a diagnostic tool
for high energy  laser-matter interaction in particular for warm dense
matter \cite{Glenzer03,Hoell04,Glenzer07,Thiele08}. 
The Thomson signal probes the dynamic structure factor of the
plasma \cite{Chihara87}. 
Reversing the argument, we can synthesize the Thomson signal
by using an appropriate expression for the dynamic structure factor
and infer density and temperature conditions by matching the
synthesized signal to the experimental one.
In particular, at high densities, collisions and correlations 
have to be accounted for in modeling $S(k,\omega)$ \cite{Ichimaru82}.
The Mermin approximation \cite{Mermin} has been found to be a simple way of
interpolating between the collision-less plasma (RPA)
at large wavevectors and collisions in the long wavelength 
limit, i.e. a Drude-like expression for the dielectric function. 
However, in non-ideal plasmas, correlations beyond the RPA 
exist even at finite values of $k$ and have to be taken into 
account. The traditional Mermin approach fails to incorporate these
correlations. 
An interpolation scheme between static local field corrections and the
Drude model by a generalized Mermin approach has been suggested in Ref.
\cite{AW} using the Zubarev approach to the non-equilibrium
statistical operator \cite{Zubarev97,Mermin_Zubarev}. 
This scheme guarantees the correct account of electron
correlations in the static limit.

Having an interpolation scheme at our disposal, a systematic study of
the influence of both electron-ion collisions as well as
electron-electron correlations is possible. It is the objective of this
paper to contribute to such a study. In particular, since the correct
form of the dynamic as well as  the static local field corrections
for the interacting electron gas is still a matter of debate, we 
compare a few recent suggestions in their consequences for the Thomson
scattering signal.

The paper is organized as follows. In Sec.~\ref{sec:Background}, we
give a brief review of the formalism and the approach to the dynamic
collision frequency. 
Sec.~\ref{sec:local_fields} in some detail explains the
different models for the dynamic local field correction considered
here. Results for the plasmon dispersion are discussed in
Sec.~\ref{sec:numerical_comparison}. Finally, conclusions and an
outlook complete this paper.

\section{Theoretical background}
\label{sec:Background}

We consider a neutral plasma of electrons and ions in thermal
equilibrium with electron density $n_e$, ion density $n_i=n_e$ and
temperature $T$. For later use, we introduce the Fermi wave vector
$k_F=(3 \pi^2 n_e)^{1/3}$, the Fermi energy $E_f=\hbar^2 k_F^2/(2
m_e)$, and the Brueckner parameter $r_s$ given by
$(4 \pi/3) n_e a_{\rm B}^3 r_s^3\,=\,1$, where $a_{\rm B}$ is 
Bohr's radius. These parameters are relevant in our context, because 
we use the model of an electron gas at $T=0$ interacting with an 
inert background of ions in carrying out our exploratory calculations.

\subsection{Thomson scattering and Born-Mermin approach}
\label{secsec:Background_Thomson}

It is well-known, see
\cite{Hoell04,Glenzer07,Thiele08}, that 
the experimental Thomson scattering cross section is
related to the dynamic structure factor of all electrons in the
plasma according to
\begin{equation}
\label{Chihara_1} \frac{d^2\sigma}{d\Omega
d\omega}=\sigma_T\frac{k_1}{k_0}S_{ee}(k,\omega )\,\,.
\end{equation}
In this expression, 
$\sigma_T=6.65\times10^{-29}\,{\rm m}^2$ is the Thomson cross
section, and $k_0 $ and $k_1$ are the wavenumbers of the incident and the
scattered light. The energy and momentum transfer are given by
$\Delta E=\hbar\omega=\hbar\omega_1-\hbar \omega_0$ and
$\hbar\mathbf{k}=\hbar\mathbf{k}_1-\hbar\mathbf{k}_0$. The momentum
is related to the scattering angle $\theta_s$ in the limit
$\hbar\omega\ll\hbar\omega_0$ by
$k=4\pi\sin(\theta_s/2)/\lambda_0$ for an
 incident wavelength $\lambda_0$. Here, we follow
Chihara' s approach \cite{Chihara87}, in
that the total dynamic structure factor can be written in terms of
contributions from free electrons , weakly and tightly bound
electrons, and core electrons. In this paper, only the dynamic structure
factor of free electrons is considered.

In thermodynamic equilibrium, the dynamic structure factor
$S_{ee}(k,\omega)$ and the longitudinal response function
$\chi_{ee} (k,\omega)$ are related via the
fluctuation-dissipation theorem
\begin{eqnarray}
  \label{eq:flucatuation_dissipation}
  S_{ee}(k,\omega) & = & -\frac{1}{\pi n_e}\,\frac{1}{1-{\rm
      e}^{-\hbar \omega/(k_B T}}\,
 \mbox{Im}\, \chi_{ee}(k,\omega) \,\,\,.  
\end{eqnarray}
Theoretical approaches to dynamic structure factor of two-component 
plasmas have been developed starting from different approaches such as
perturbation theory, the viscoelastic model \cite{Ichimaru85}, 
the recurrence relation method \cite{Daligault03}, or the moment
approach, see Ref.~\cite{Adamyan93}. As an example for a perturbative
treatment, we mention Ref.~\cite{Roepke99,Reinholz}. 
There, based on the generalized linear
response theory of Zubarev, a systematic account of correlations as 
well as collisions has been accomplished by partial summation of 
diagram classes using thermodynamic Green's functions. 
While a detailed evaluation of the resulting expressions for the 
response functions is cumbersome at arbitrary wave vectors $k$, 
numerical calculations have been carried out in the long-wavelength
limit $k \to 0$. In particular, approximative expressions for the
collision frequency $\nu(\omega)$ have been studied taking care of 
strong collisions as well as dynamical screening in a consistent
manner. 

To generate approximative results for the response function at
finite wave vectors $k$, we follow an idea suggested by
Mermin~\cite{Mermin}. 
Ensuring particle number conservation by introducing local thermal
equilibrium together with a relaxation time ansatz, the electron-electron
response function is approximated by
\begin{eqnarray}
  \label{eq:trad_mermin}
    \chi^{(M)}(k,\omega) & = & 
  \left( 1- \frac{i \omega}{\eta} \right) 
  \frac{\chi_{\rm ee}^{\rm RPA}(k,\omega+i\eta)\,\chi_{\rm ee}^{\rm RPA}(k,0)}{
        \chi_{\rm ee}^{\rm RPA}(k,\omega+i\eta)-\frac{i \omega}{\eta} 
        \chi_{\rm ee}^{\rm RPA}(k,0)} \,\,\,,
\end{eqnarray}
with a relaxation parameter $\eta$. For details, see Ref.~\cite{Mermin_Zubarev}.
Here, $\chi_{\rm ee}^{\rm RPA}$ is the electron response function in
random phase approximation, i.e.
\begin{eqnarray}
  \label{eq:chi_RPA}
  \chi_{\rm ee}^{\rm RPA}(k,\omega) & = & 
  \frac{\chi^{(0)}(k,\omega)}{1\,-\,V(k) \chi^{(0)}(k,\omega)}
\end{eqnarray}
where $\chi^{(0)}(k,\omega)$ is the ideal, i.e non-interacting 
response function ,
see \cite{Arista84}.
For $T=0$, this ideal response can be found as
\begin{eqnarray}
  \label{eq:ideal_response_T=0}
  V(k) \chi^{(0)}(k,\omega) & = & 
 - \frac{\chi_0^2}{4 z^3} \,
  \left[ g(u+z)-g(u-z) \right]\,\,\,,
\end{eqnarray}
with $u\,=\,m \omega/(\hbar k\, k_F)\,,\,z  \, = \, k/(2
  \,k_F)\,,\,\chi_0^2 \, = \, \left( \pi k_F a_{\rm B}\right)^{-1}$ and 
\begin{eqnarray}
  \label{eq:ideal_response_definitions}
  g(x) & = & x+ \frac{1}{2} \left( 1-x^2 \right) \,\mbox{ln} 
              \frac{x+1}{x-1}  \,\,\,, \\
\end{eqnarray}
The function $g(x)$ given here is a generalization of the function $g(x)$ 
given in \cite{Arista84} to complex arguments.
Furthermore $V(k)$ denotes the Coulomb-potential in momentum space.
Note, that Eq.~(\ref{eq:trad_mermin}) 
reduces to the RPA expression in the absence 
of collisions, i.e. $\eta=0$. Also, in the long-wavelength limit, 
Eq.~(\ref{eq:trad_mermin}) turns into the familiar Debye form, allowing 
to identify the relaxation parameter $\eta$ as the 
collision frequency $\nu(\omega)$.

\subsection{Extended Mermin approach}
\label{secsec:Background_extended_Mermin}

To account for correlations among the electrons, an extension of 
the traditional Mermin expression has been suggested, 
see Ref.~\cite{AW}. For an adiabatic model with inert ions, 
it reduces to replacing the RPA response function in
Eq.~(\ref{eq:trad_mermin}) by the response function of 
the interacting one-component (OCP) electron gas $\chi^{\rm OCP}_{ee}(k,\omega)$,
\begin{eqnarray}
  \label{eq:extended_Mermin}
  \chi^{(\rm xM)}_{ee}(k,\omega) & = & \left( 1- \frac{i \omega}{\eta}
\right) \, \left( \frac{ \chi^{\rm OCP}_{ee}(k, \omega+i
\eta)\,\chi^{\rm OCP}_{ee}(k,0)}{ \chi^{\rm OCP}_{ee}(k,\omega+i \eta)\,-\,
\left(i \omega /\eta \right) \chi^{\rm OCP}_{ee}(k,0)} \right) \,\,\,,
\end{eqnarray}
where the label xM indicates the extended Mermin expression for the
response function. Note, that the same expression has been derived 
independently by Barriga-Carrasco~\cite{Barriga-Carrasco09}.
Traditionally, the OCP 
response function is represented using a
dynamic local field correction $G_{ee}(k,\omega)$ as 
\begin{eqnarray}
   \chi^{\rm OCP}_{ee}(k,\omega) & = & 
   \frac{\chi^{(0)}_e(k,\omega)}{1-V(k) \left(
   1-G_{ee}(k,\omega) \right) \chi^{(0)}_e(k,\omega)}\,\,\,.
\end{eqnarray}
Having a collision-less plasma $\nu=0$, the response function is solely the
OCP expression. Due to the fact, that electron-electron collisions do not 
contribute in the long-wavelength limit, i.e. 
$G_{\rm ee}(k,\omega) \propto k^2$ for $k \to 0$, this 
expression still reduces to the Drude-like form for small $k$
with the same $\nu(\omega)$ as before. However, contrary to 
Eq.~(\ref{eq:trad_mermin}), the static limit is now given 
by the electron-electron correlation in the OCP,
\begin{eqnarray}
  \label{eq:extended_Mermin_static_limit}
   \lim_{\omega \to 0}   \chi^{(\rm xM)}_{ee}(k,\omega) & = & 
   \chi^{\rm OCP}_{ee}(k,0)\,\,\,.
\end{eqnarray}

\subsection{Collision frequency at arbitrary degeneracy}
\label{secsec:Background_collision_frequency}

\begin{figure}
\includegraphics[width=12cm]{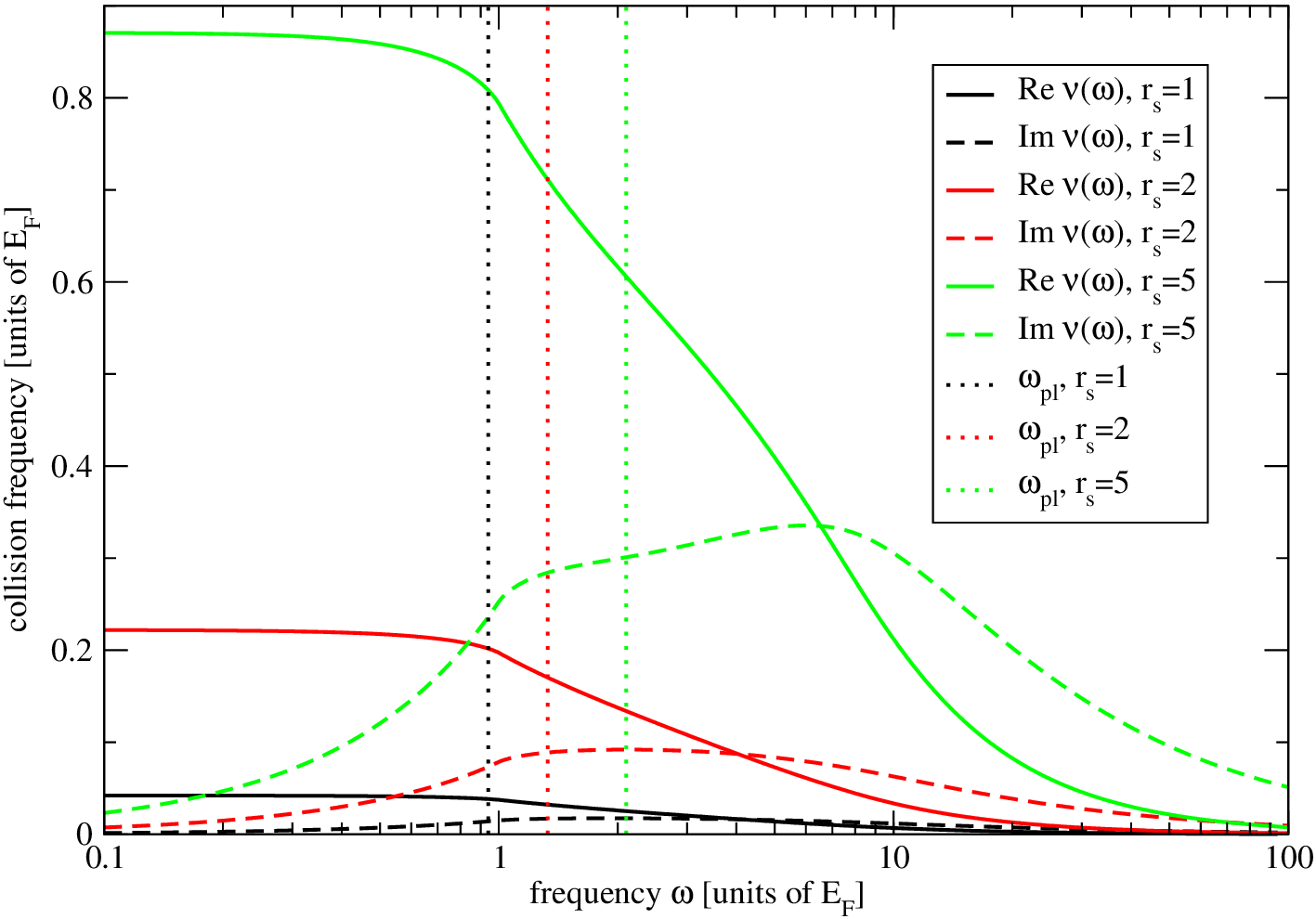}
\caption{Real part of the collision frequency $\nu(\omega)$ as a
 function of the frequency $\omega$. Various values of the Brueckner 
 parameter $r_s$ are considered.  }
\label{fig:coll_freq_born_dynamic}
\end{figure}

For the exploratory calculation discussed here, we use the collision
frequency in Born approximation and for arbitrary degeneracy, 
see Ref.~\cite{Reinholz},
\begin{eqnarray}
  \label{eq:nu_omega}
  \mbox{Re}\,\nu(\omega) & = & 
  \frac{\epsilon_0 n_i \Omega_0^2}{6 \pi^2 e^2 n_e m_e} \,
  \int_0^{\infty} \!dq\,q^6 \,V_{\rm TF}(q)^2\,S_i(q)\,\frac{1}{\omega}
  \mbox{Im}\,\epsilon_{\rm RPA,e}(q,\omega) \,\,\,,
\end{eqnarray}
where $V_{\rm TF}(q)=V(q)/\epsilon^{(RPA)}(q,0)$ 
is the static screened potential, $S_i(q)$ is the 
static structure factor of the ions, taken e.g. in HNC approximation
or from MD simulations, $\Omega_0$ is a normalization volume, and 
$\epsilon_{\rm RPA,e}(q,\omega)$ is the dielectric function of the 
electron OCP for arbitrary degeneracy as given e.g. in the article 
of Arista and Brandt \cite{Arista84}. Again, we determine the RPA
dielectric function for $T=0$ by using Eq.~(\ref{eq:ideal_response_T=0}).

We restrict
ourselves to this easily accessible expression since we want to focus on
the interplay between collisions and electron-electron correlations
and the r\^{o}le of different approximations for the OCP local field corrections. 
More advanced expressions are given in Ref.~\cite{Reinholz} and should
be used for realistic calculations. Also, realistic calculations
should include electron-electron effects on the collision frequency,
which can be taken into account by increasing the number of moments 
in linear response approach, see Ref.~\cite{Reinholz} as well.

In Fig.~\ref{fig:coll_freq_born_dynamic}, we show the 
collision frequency $\nu(\omega)$ as a function of the frequency 
$\omega$ for three different values of the Brueckner parameter 
$r_s=1,2,5$. Note, that $\nu(\omega)$ is a complex quantity and the
imaginary part is connected to $\mbox{Re}\,\nu(\omega)$ by a
Kramers-Kronig relation
\begin{eqnarray}
  \label{eq:im_nu_Kramers_Kronig}
  \mbox{Im}\,\nu(\omega) & = & 
  \int_{-\infty}^{\infty}\!\frac{d
    \omega'}{\pi}\,\frac{\mbox{Re}\,\nu(\omega')}{
  \omega-\omega'}\,\,\,.
\end{eqnarray}
The account of this imaginary part is essential for obeying both, the
f-sum sum rule and the perfect screening sum rule. 
While loosely speaking, the real part leads to a broadening of the 
plasmon at $k=0$, the imaginary part produces a shift of the
plasmon. For a static frequency $\omega=0$, the imaginary part
vanishes, i.e. replacing the dynamic by  a static collision frequency
$\nu(0)$ one ignores the shift of the plasmon position.

\section{Local field corrections for an interacting electron gas at
  $T=0$}
\label{sec:local_fields}

\subsection{Static local field correction for the OCP}
\label{secsec:Gkw_static}

\begin{figure}[t]
\includegraphics[width=12cm]{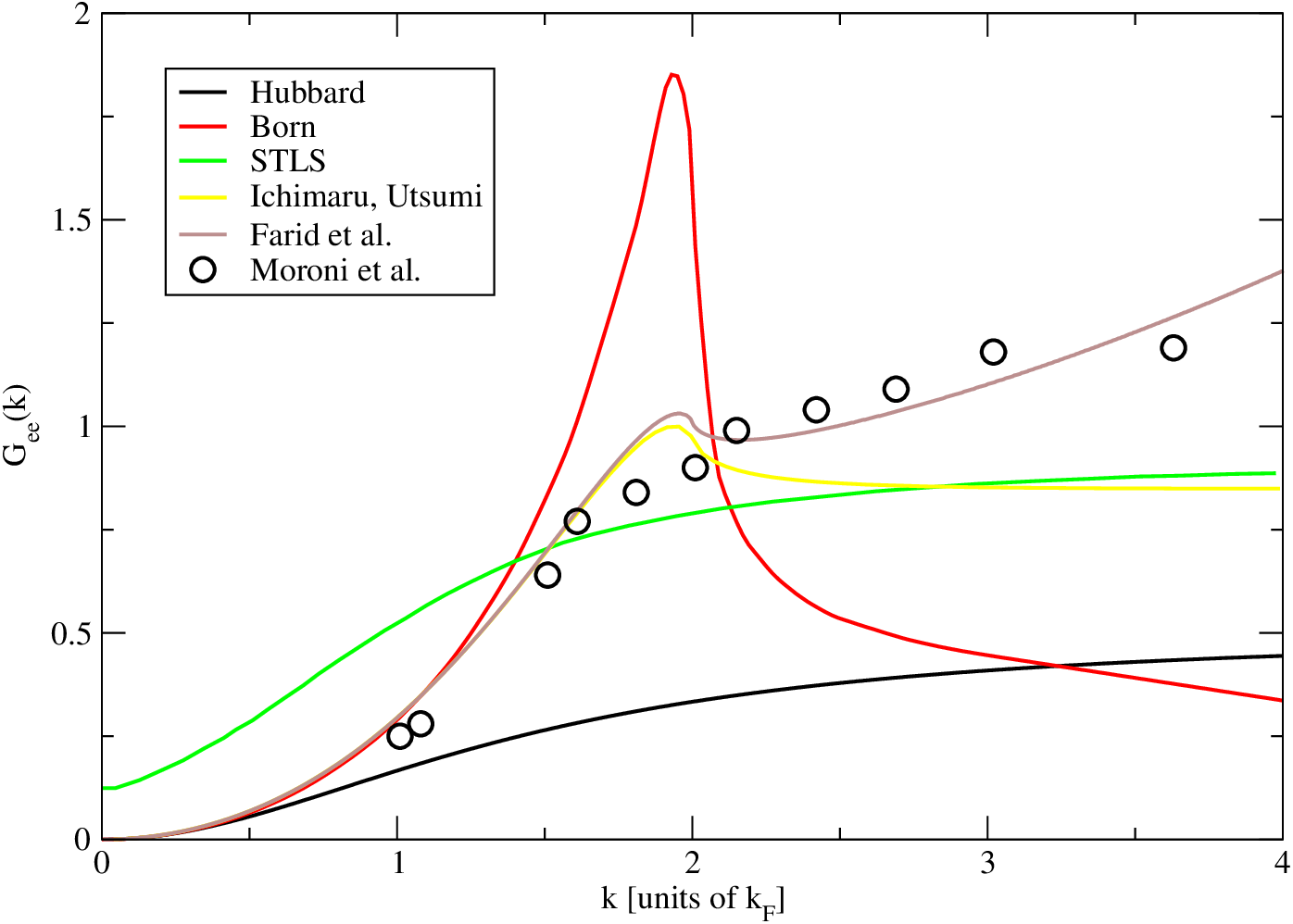}
\caption{Static local field correction $G_{\rm
    ee}(k)$ for $r_s=2$. Approximations: Hubbard~\cite{Hubbard57},
  Born~\cite{Engel90}, STLS~\cite{STLS}, Utsumi and
  Ichimaru~\cite{Utsumi80},
  Farid et al.~\cite{Farid93}, Moroni et al.~\cite{Moroni95}.}
\label{fig:static_local_fields}
\end{figure}

In an often used approximation, the dynamics in the local field
correction is ignored reducing it to the static limit only,
\begin{eqnarray}
  \label{eq:chi_static_local_field}
   \chi^{\rm OCP}_{\rm ee}(k,\omega) & = & 
   \frac{\chi^{(0)}_e(k,\omega)}{1-V(k) \left(
   1-G_{\rm ee}(k) \right) \chi^{(0)}_e(k,\omega)}\,\,\,.
\end{eqnarray}
For the static local field correction $G_{\rm ee}(k)$, a plethora
of approximations have been suggested beginning with the original
paper by Hubbard~\cite{Hubbard57}. Here, it is impossible to give an 
exhaustive review.
Instead, we concentrate to the widely-used expression of 
Ichimaru and Utsumi~\cite{Utsumi80} 
and its extension by Farid et al.~\cite{Farid93}. 
For an overview of other approximations, see Ref.~\cite{Ichimaru94}.
For the sake of illustration, we compare a few approximations in 
Fig.~\ref{fig:static_local_fields} for $r_s=2$, see \cite{Engel90,STLS}. 
Also included are results obtained with a Monte
Carlo (MC) simulation by Moroni et al.~\cite{Moroni95}.
Qualitatively similar results 
are obtained for other values of $r_s$. For small values of $k$,
the Utsumi-Ichimaru approximation and the extended model of Farid et
al. are identical by construction and adapted to the compressibility sum
rule. There is a good agreement with the MC data and both, 
the Utsumi-Ichimaru and the Farid et al. description, 
while the other approximations do not describe these so well. 
In the deep inelastic regime at large $k$, Farid et al. take account of the 
results by Holas~\cite{Holas}, that $G_{\rm ee}(k)$ scales as $k^2$, which
is not included into the Utsumi-Ichimaru ansatz. The MC data
seem to support this $k^2$ scaling. For our discussion, this
difference is rather unimportant, since the plasmon ceases to be a 
well-defined mode at $k_0 \le k_F$, while the differences between
Ichimaru-Utsumi and Farid et al. arise for $k \gtrsim 2 k_F$.

\subsection{Dynamic local field corrections for the OCP}
\label{secsec:Gkw_dynamic}

As already mentioned, there are several approximative approaches to the 
dynamic structure factor of the electron gas. Here, we use the
approach of Dabrowski \cite{Dabrowski86} and the approach of Hong and
Lee \cite{Hong85}. Both approaches are interpolation schemes incorporating
sum rules and other exact properties. In particular, the static
properties are inputs into these schemes and we can use the local
field correction of Farid et al. again in this case. This would have
been impossible if we choose the perturbative results of
e.g. Richardson and Ashcroft~\cite{Richardson94}.

Being a dynamical quantity, the local field correction is also a
complex quantity obeying Kramers-Kronig like relations. 
For the real part $\mbox{Re} \,G_{\rm ee}(k,\omega)$ of the dynamic local field 
correction, the static limit is just approximated by $G_{\rm ee}(k)$ given
above. The high frequency asymptotics is given by three-frequency sum
rule and therefore by the static structure factor and 
the correlated kinetic energy of the electron gas,
see Ref.~\cite{Iwamoto84,Lee89},
\begin{eqnarray}
  \label{eq:G_k_w_high_frequency}
  \lim_{\omega \to \infty}  G_{\rm ee}(k,\omega) & = & 
  I(k)\,-\,\frac{2 k^2}{m \omega_{\rm pl}^2}
  \left( \langle E_{\rm kin} \rangle \,-\, \langle E_{\rm kin}
    \rangle_0  
  \right) \,\,\,
\end{eqnarray}
which in turn is given by the static structure factor $S(k)$
\begin{eqnarray}
  \label{eq:I_k_S_k}
  I(k) & = & - \frac{1}{N} \sum_{\vec q \neq \vec k, \vec 0} 
  K\left(\vec k,\vec q\right)\left( \vec k \cdot \vec q \right)^2
  \left[ S(|\vec q -\vec k|)\,-\,1 \right]\,\,\,,
\end{eqnarray}
and 
\begin{eqnarray}
  \label{eq:I_k_K_k}
  K(\vec k,\vec q) & = & 
  \frac{\vec q \cdot \vec k}{k^2}\,+\,\frac{\vec q \cdot
  \left( \vec q -\vec k \right)}{|\vec q-\vec k|^2 }\,\,\,.
\end{eqnarray}
Furthermore, $\langle E_{\rm kin}\rangle $ is the kinetic energy of
the interacting electron gas, $ \langle E_{\rm kin}\rangle_0 $ its
non-interacting counterpart, $\omega_{\rm pl}$ is the plasma frequency.

Also, Dabrowski incorporates the perturbative result of Glick and 
Long~\cite{Glick71} for the high-frequency behavior of the imaginary 
part of the dynamic structure factor.  He extends a Pad\'{e}
approximation suggested by Gross and Kohn~\cite{Gross85} to 
finite values of the wave vector, 
\begin{eqnarray}
  \label{eq:Im_G_ansatz}
  \mbox{Im}\, G_{\rm ee}(k,\omega) = \frac{ a(k) \omega}{(1+b(k) \omega^2)^{5/4}}
\end{eqnarray}
with
\begin{eqnarray}
  \label{eq:a_k_b_k}
  a(k) & = & C k^2 \left( \frac{\mbox{Re}\,G_{\rm ee}(k,0)\,-\,
  \mbox{Re} G_{\rm ee}(k,\infty)}{C D k^2} \right)^{5/3} \,\,\,,\\
  b(k) & = & \left( \frac{\mbox{Re}\,G_{\rm ee}(k,0)\,-\,
      \mbox{Re}\,G_{\rm ee}(k,\infty)}{C D k^2} \right)^{4/3} \,\,\,,
\end{eqnarray}
and $C=23/66\, \alpha\, r_s$, $D=\Gamma(3/4)/(\sqrt{\pi} \,
\Gamma(5/4)) \approx 0.763 $, $\alpha=(4/(9 \pi))^{1/3}$ .
The corresponding real part is then obtained by a Kramers-Kronig
relation,
\begin{eqnarray}
  \label{eq:Re_G_ansatz}
  \mbox{Re}\,G_{\rm ee}(k,\omega) & = &
  \mbox{Re}\,G_{\rm ee}(k,\infty) \,+\,
  P\int_{-\infty}^{\infty} \frac{d \omega'}{\pi}\,
  \frac{\mbox{Im}\,G_{\rm ee}(k,\omega')}{\omega'\,-\,\omega} \,\,\,, 
\end{eqnarray}
where $P\int$ indicates Cauchy principal value integration.

As a second option for the dynamic local field correction of the OCP
we introduce the above mentioned approach of Hong
and Lee~\cite{Hong85,Lee89}, which is based on the recurrence relation
technique. Specifically, we use the lowest dynamical extension 
of the local field correction, which can be introduced by this
technique. Adapting the notation to this paper, the dynamical local field 
correction reads
\begin{eqnarray}
  \label{eq:Gkz_recurrence_relation}
  G_{\rm ee}(k,z) & = & 
  G_{\rm ee}(k,0) \,+\, \left[ G_{\rm ee}(k,\infty) \,-\,
  G_{\rm ee}(k,0)\right]\,c_2^0(z)
\end{eqnarray}
with a function $c_2^0$ given by~\cite{Anmerkung}
\begin{eqnarray}
  \label{eq:c_2^0}
  c_2^0(z) & = & \frac{\Delta_1^0}{\Delta_2^0} \left( 
  \frac{ \chi^{(0)}(k,0)}{ \chi^{(0)}(k,z)} \,-\,1 \right) 
  + \frac{z^2}{\Delta_2^0}\,\,\,.
\end{eqnarray}
Here, the quantities $\Delta_1^0$ and $\Delta_2^0$ are the ideal
recurrants $\Delta_1^0=-\frac{\omega_{\rm pl^2}}{V(k) \chi_{\rm
    ee}^{(0)}(k)}\,\,,\,\,
\Delta_2^0=\left[\frac{12}{5} \left(\frac{k}{k_F}\right)^2 +
\left(\frac{k}{k_F}\right)^4 \right] \left(\frac{E_F}{\hbar}
\right)^2-\Delta_1^0$.

Both approaches have been adapted in this paper to the most recent 
results for the static electron-electron structure factor $S(k)$
and the correlated kinetic energy, see Ref.~\cite{Gori-Giorgi00}.
The details will be covered in an forthcoming publication \cite{Wierling09}.

\section{Comparison of different approximations}
\label{sec:numerical_comparison}

\begin{figure}[t]
\includegraphics[width=12cm]{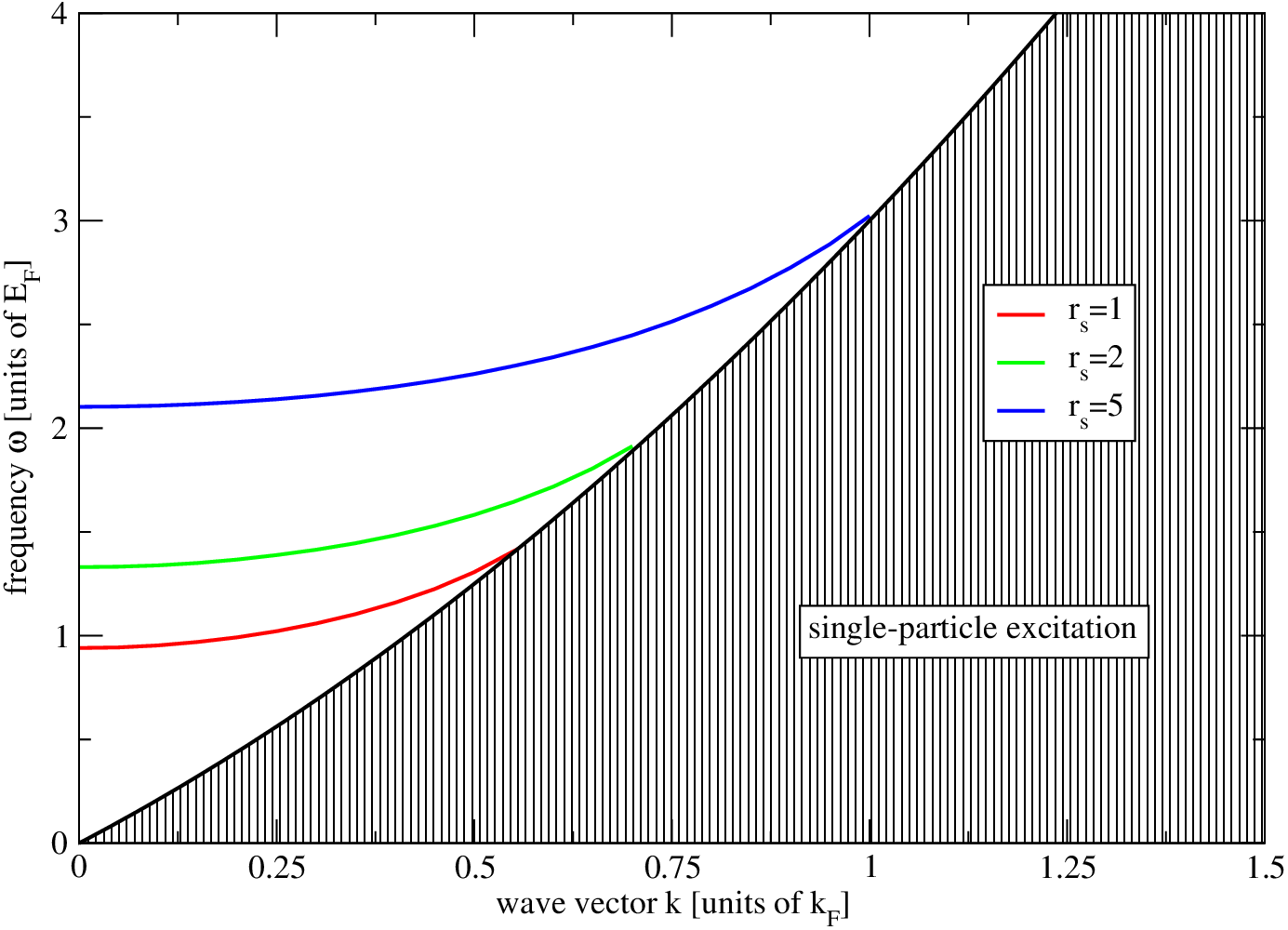}
\caption{Plasmon dispersion relation in RPA approximation. 
 Brueckner parameters $r_s=1,2,5$ are considered. The intersection
 of the plasmon dispersion with the single-particle ridge defines
 the wave vector $k_0(r_s)$.}
\label{fig:dispers_RPA}
\end{figure}

\begin{figure}[t]
\includegraphics[width=12cm]{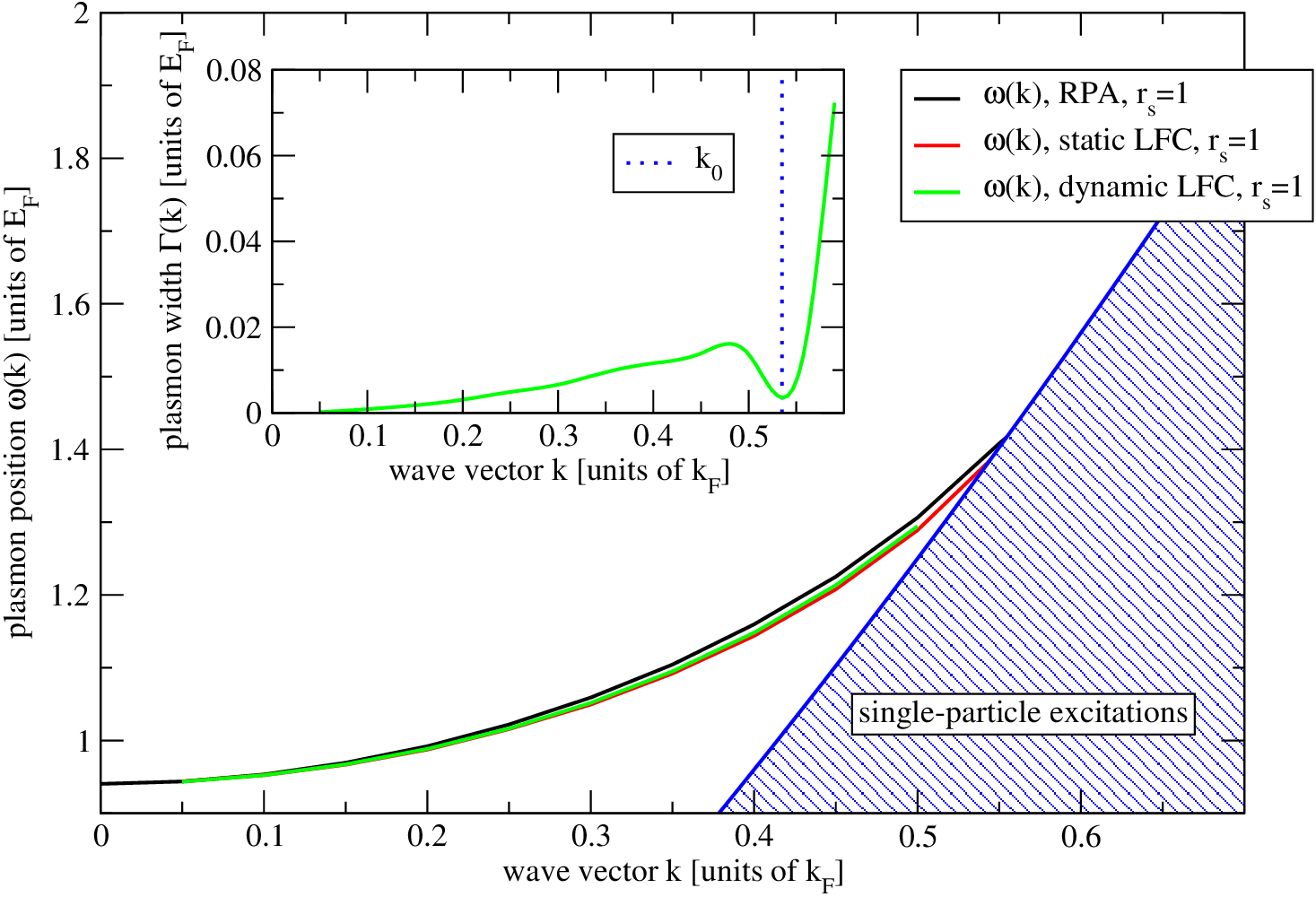}
\caption{Plasmon shift $\omega(k)$ and plasmon width $\Gamma(k)$
as a function of the wave vector k. $r_s=1$ is studied. 
We compare the static local field correction to the dynamic local
field correction. No collisions are included.
}
\label{fig:dispersion_dynamic_LFC_rs=1}
\end{figure}

Instead of showing the dynamical structure factor $S(k,\omega)$ for 
each of the different approximations and for various values of the 
wave vector $k$ and the frequency $\omega$, we introduce the plasmon
position $\omega(k)=\underset{\omega}{\rm max}\, S_{\rm ee}(k,\omega)$ 
and the plasmon width $\Gamma(q)$ 
as a signature for the influence of different 
effects, where we measure the width as full width at half maximum (FWHM).
In a collision-less plasma at $T=0$, the plasmon is a well-defined mode
for $k\,<\,k_0(r_s)$ corresponding to a $\delta$-like spike
in the dynamic structure factor. For larger $k$, the plasmon 
ceases to be a well-defined mode and shows a broadening even in RPA, 
see Ref.~\cite{Arista84}. We illustrate these dispersion relations in 
Fig.~\ref{fig:dispers_RPA}. The wave vector $k_0(r_s)$, where the 
plasmon dispersion $V(k) \chi^{(0)}(k,\omega)\,=\,1$ intersects with
the
single-particle ridge $\omega=\hbar k \left(k-2 k_F \right)/2 m$ is
smaller than $k_F$ for all of the values $r_s$ considered here.
Collisions as well as correlations 
modify these RPA dispersion relations leading to a shift and a broadening
of the plasmon even at $k\,<\,k_0(r_s)$.
For the conditions of warm dense matter, the
plasmon dispersion in the traditional Born-Mermin approach has been 
studied by Thiele et al. \cite{Thiele08}.

\subsection{Plasmon dispersion without collisions}
\label{secsec:without_collisions}

\begin{figure}[t]
\includegraphics[width=12cm]{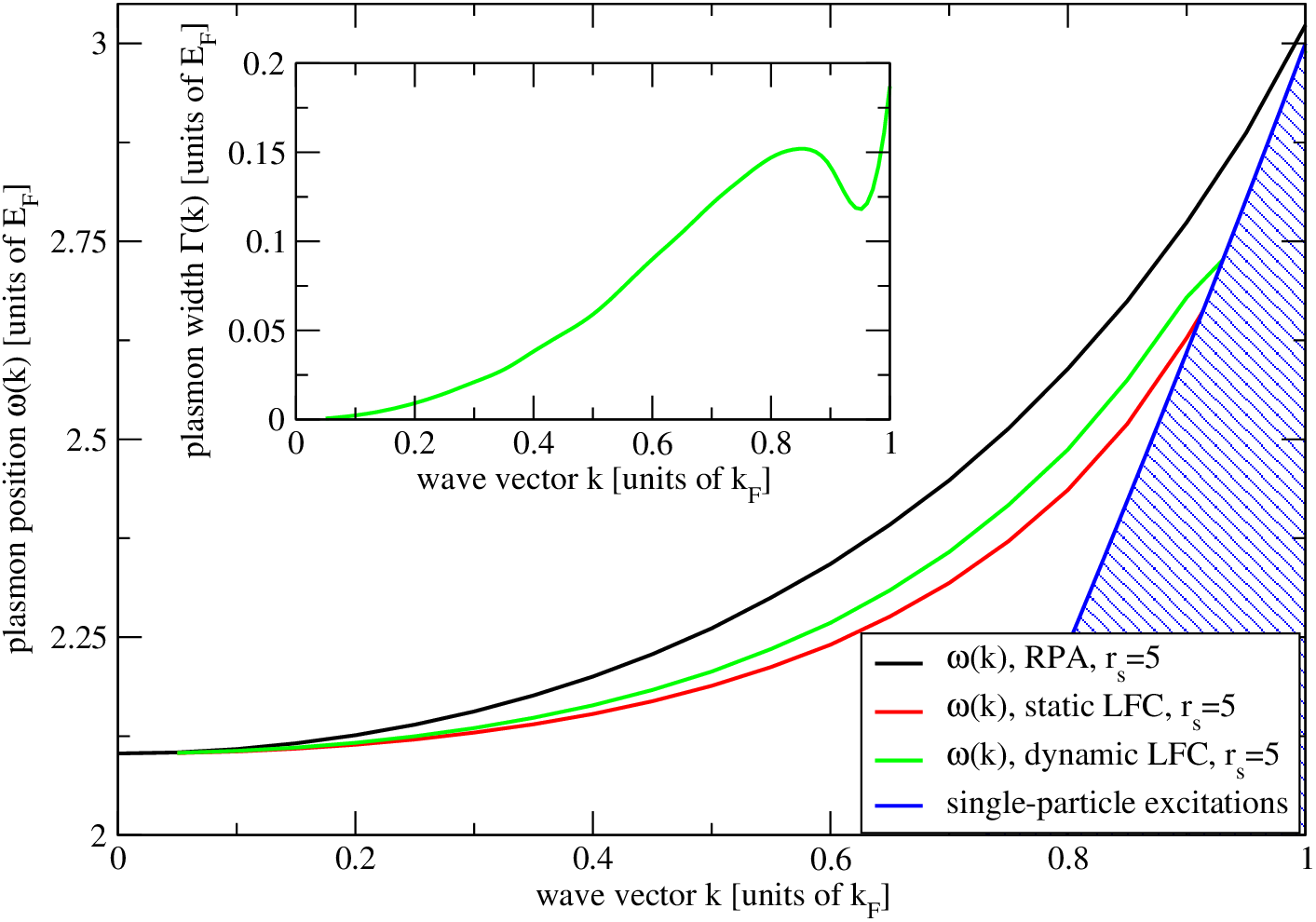}
\caption{Plasmon shift $\omega(k)$ and plasmon width $\Gamma(k)$
as a function of the wave vector k. $r_s=5$ is studied. 
We compare the static local field correction to the dynamic local
field correction. No collisions are included.
}
\label{fig:dispersion_dynamic_LFC_rs=5}
\end{figure}

In a first step, we discuss the effects induced only by
local field corrections, i.e. for $\nu(\omega)=0$.
The modifications of the plasmon properties are shown in 
Fig.~\ref{fig:dispersion_dynamic_LFC_rs=1} for $r_s=1$ and 
in Fig.~\ref{fig:dispersion_dynamic_LFC_rs=5} for $r_s=5$.
RPA, static local field corrections and dynamic local fields given by
the improved Dabrowski interpolation scheme are compared.
In the first case, the deviations of the static LFC 
from the RPA results are quite small. Also, the dynamic LFC is only a
minor correction to the static expression. This is consistent with the 
plasmon width, which is below 1 \% for almost all wave vectors
$k$. Note, that the rapid increase above $0.55\,k_F$ is due to the
onset of damping in the RPA expression. For $r_s=5$, the noticeable deviations 
occur. Also, a clear influence of dynamical LFC is visible. This is
also reflected in the plasmon damping, where a width of up to 17 \% is
found.

\subsection{Results for the extended Mermin approach}
\label{secsec:plasmons_extended_Mermin}

\begin{figure}[t]
\includegraphics[width=12cm]{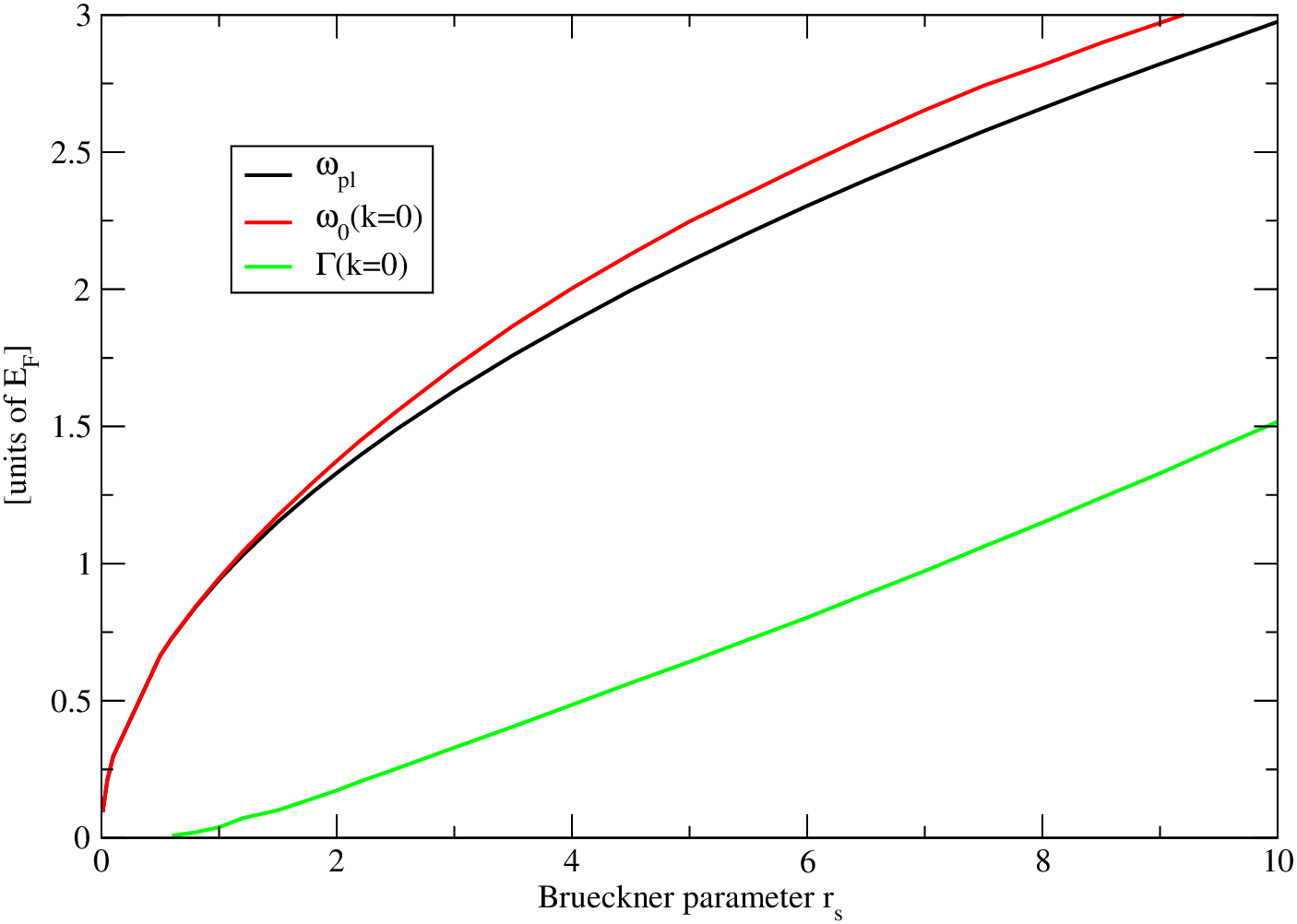}
\caption{Plasmon shift $\omega(0)$ and width $\Gamma(0)$  
as a function of the Brueckner parameter $r_s$. 
Here, we account for electron-ion collisions only using the
traditional Born-Mermin approximation of Eq.~(\ref{eq:trad_mermin})
and Eq.~(\ref{eq:nu_omega}). No local field corrections are considered.
}
\label{fig:plasmon_k=0_nu}
\end{figure}

We start the presentation of the results for the extended Born-Mermin
approach by focussing on the long-wavelength limit $k \to 0$. In this
limit, the Born-Mermin ansatz reduces to a Drude-type dielectric
function with a frequency-dependent and complex collision frequency
$\nu(\omega)$. This frequency leads to a broadening and a shift of the 
plasmon as can be seen from the imaginary part of the inverse
dielectric function, which is given by 
\begin{eqnarray}
  \label{eq:Inverse_dk_Drude}
  \mbox{Im}\,\epsilon(k \to 0,\omega) & = & 
  -\,\frac{\mbox{Re}\,\nu(\omega)\,\omega\,\omega_{\rm pl}^2}{
  \left( \omega^2\,-\,\omega_{\rm
      pl}^2\,-\,\mbox{Im}\,\nu(\omega)\,\omega\right)^2
  \,+\,(\mbox{Re} \nu(\omega))^2\,\omega^2}\,\,\,.
\end{eqnarray}
Approximately, for $\mbox{Re}\, \nu(\omega) >> \mbox{Im}\, \nu(\omega)$,
the real part is connected to a broadening, while the imaginary part
induces a shift of the plasmon.
Local field corrections do not play any r\^{o}le in this
limit as discussed above. The shift and the broadening are illustrated in 
Fig.~\ref{fig:plasmon_k=0_nu} as a function of the Brueckner parameter
$r_s$. Since the real as well as the imaginary part of $\nu(\omega)$
increase with $r_s$ for the conditions considered here, the broadening
$\Gamma(k=0)$ of the plasmon also increases with $r_s$. For large
$r_s$, the broadening is about half the size of the plasma frequency
$\omega_{\rm pl}$.
The shift of the plasmon also increases with $r_s$. However, the shift
is less pronounced compared to the width of the plasmon, an effect
consistent with $\mbox{Im}\, \nu(\omega) < \mbox{Re}\,\nu(\omega)$.

\begin{figure}[t]
\includegraphics[width=12cm]{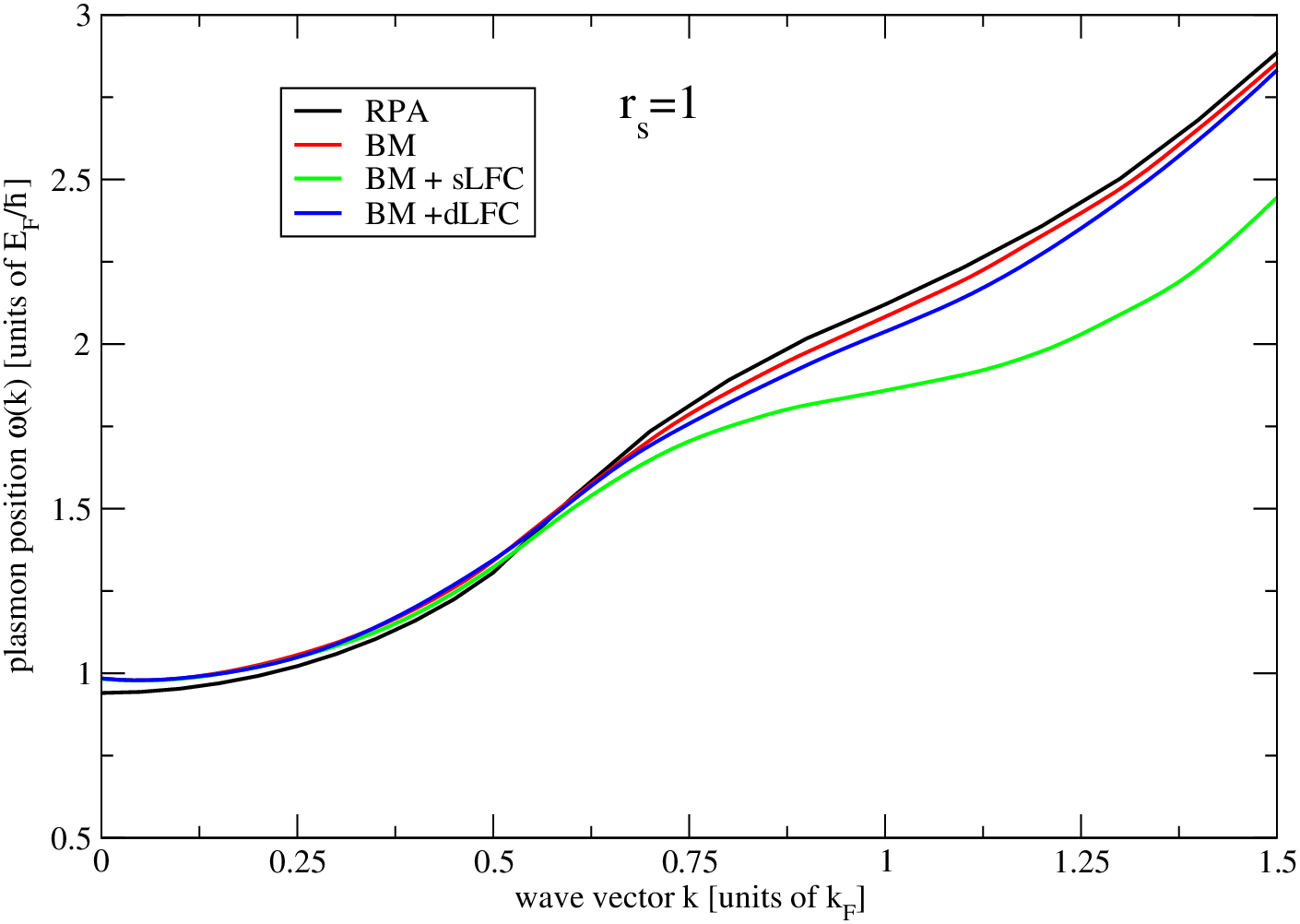}
\caption{Plasmon shift $\omega(k)$ 
as a function of the wave vector k. $r_s=1$ is studied. 
BM: traditional Born-Mermin approximation. BM + sLFC: 
Born-Mermin including static local field corrections.
BM + dLFC: Born-Mermin including dynamic local field corrections.
}
\label{fig:plasmon_shift_rs=1}
\end{figure}

\begin{figure}[t]
\includegraphics[width=12cm]{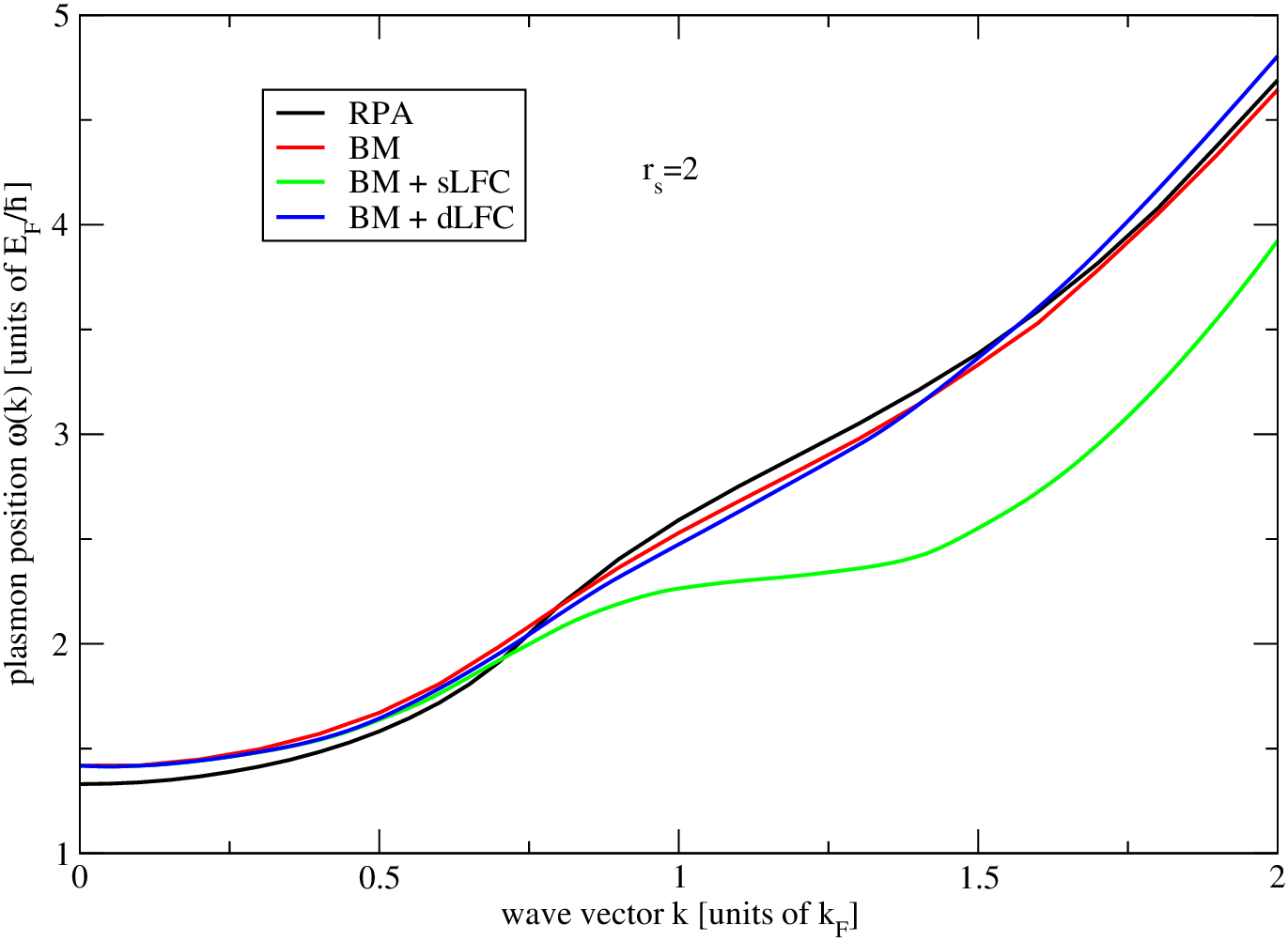}
\caption{Plasmon shift $\omega(k)$ 
as a function of the wave vector k. $r_s=2$ is studied. 
BM: traditional Born-Mermin approximation. BM + sLFC: 
Born-Mermin including static local field corrections.
BM + dLFC: Born-Mermin including dynamic local field corrections.
}
\label{fig:plasmon_shift_rs=2}
\end{figure}

\begin{figure}[t]
\includegraphics[width=12cm]{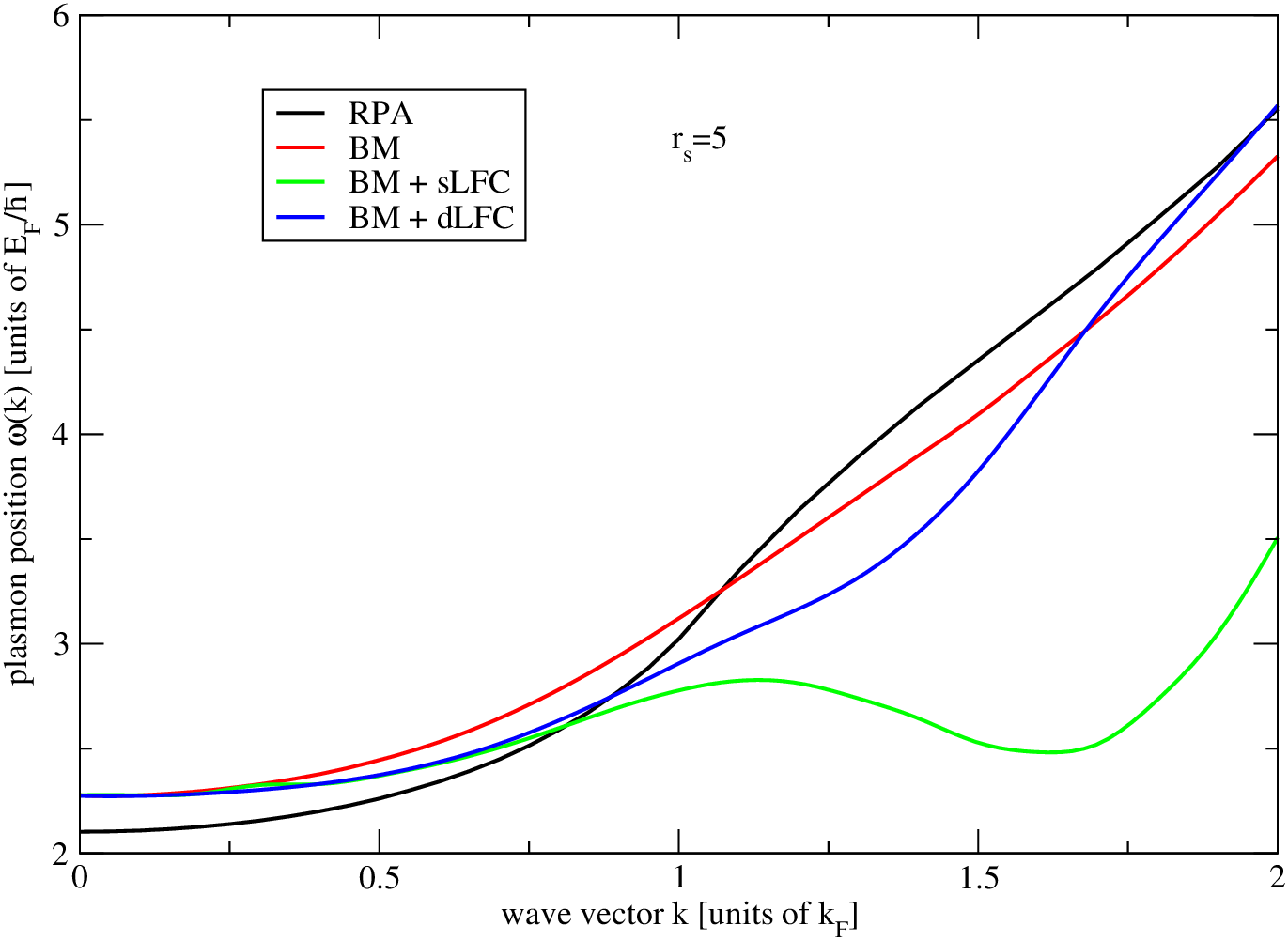}
\caption{Plasmon shift $\omega(k)$ 
as a function of the wave vector k. $r_s=5$ is studied. 
BM: traditional Born-Mermin approximation. BM + sLFC: 
Born-Mermin including static local field corrections.
BM + dLFC: Born-Mermin including dynamic local field corrections.
}
\label{fig:plasmon_shift_rs=5}
\end{figure}

\begin{figure}[t]
\includegraphics[width=12cm]{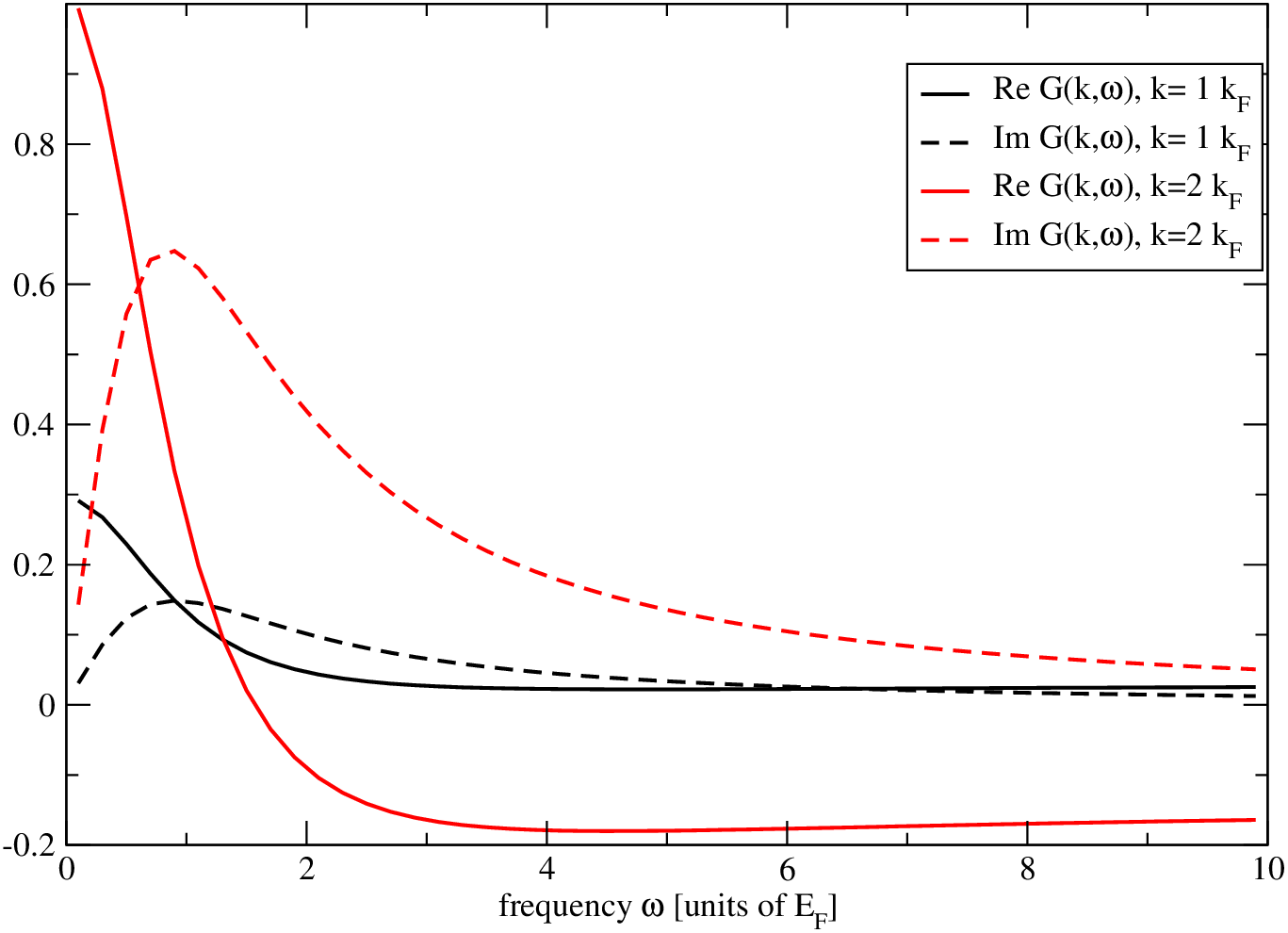}
\caption{Real and imaginary part for dynamic local field correction
at $r_s$=2 for two different wave vectors as a function of the frequency.}
\label{fig:Re_Im_G_k_omega}
\end{figure}

\begin{figure}
\includegraphics[width=12cm]{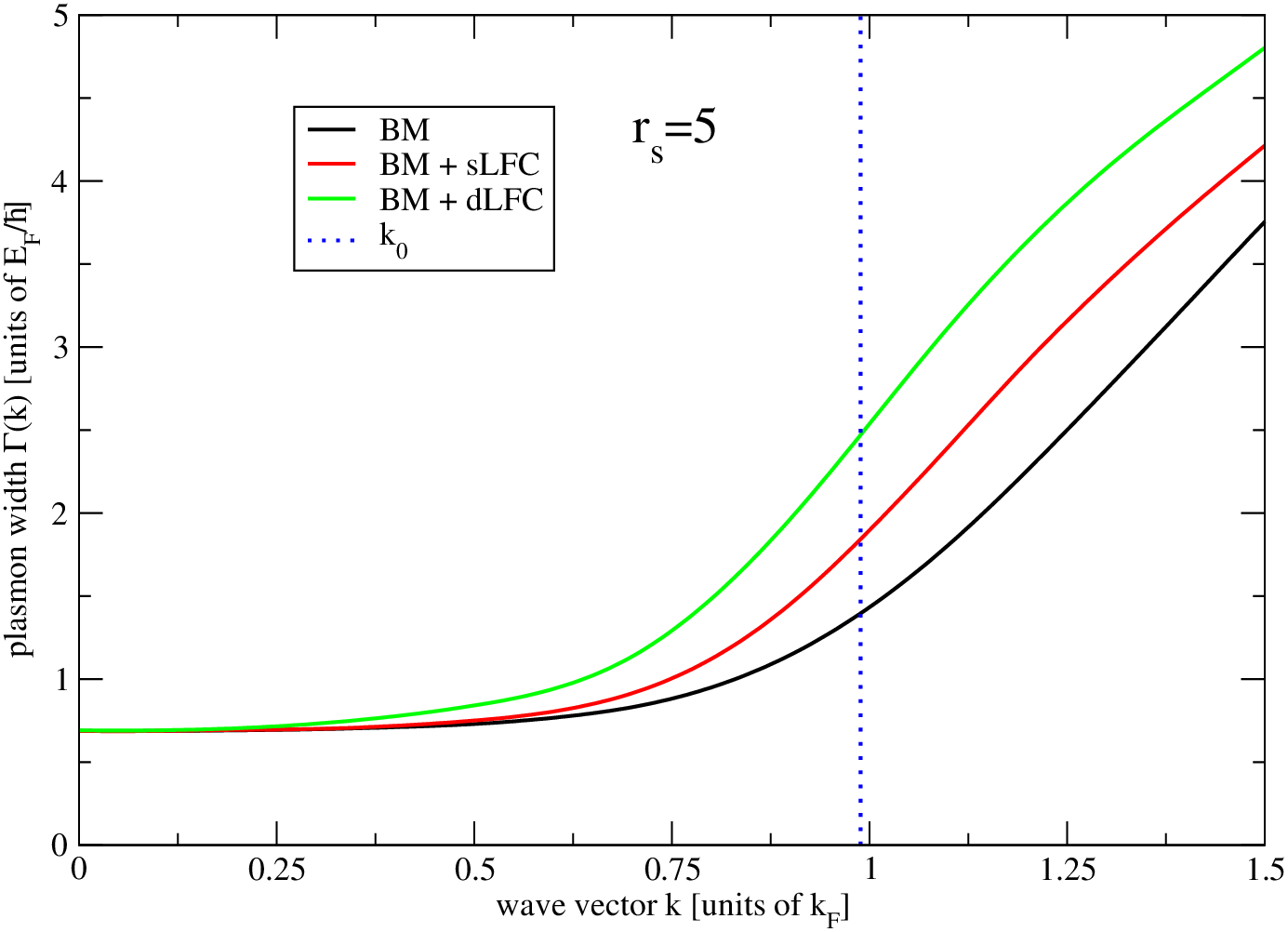}
\caption{Plasmon width $\omega(k)$ 
as a function of the wave vector k. $r_s=5$ is studied. 
The traditional Born-Mermin without local field corrections is
compared to the extended approach with static (BM + sLFC) and 
dynamic local field corrections (BM + dLFC). The onset $k_0$ of the
single-particle excitations in RPA is also shown.
}
\label{fig:plasmon_width_rs=5}
\end{figure}

Next, we present the wave vector dependence of the plasmon shift 
$\omega(k)$ and the plasmon width $\Gamma(k)$.  The results for the
shift are shown in 
Fig.~\ref{fig:plasmon_shift_rs=1}-\ref{fig:plasmon_shift_rs=5} for $r_s=1,2,5$. 
Fig.~\ref{fig:plasmon_width_rs=5} displays the width for $r_s=5$. 
We compare three different approximations, the traditional Born-Mermin
(BM) given by Eq.~(\ref{eq:trad_mermin}) and 
Eq.~(\ref{eq:nu_omega}), the extended Born-Mermin
approach with static local field correction (BM+sLFC) of 
Eq.~(\ref{eq:extended_Mermin}) together with 
Eq.~(\ref{eq:chi_static_local_field}) and finally
the extended Born-Mermin approach with dynamic local field corrections
(BM+dLFC) 
by the Dabrowski ansatz, i.e. Eq.~(\ref{eq:Im_G_ansatz}) and 
Eq.~(\ref{eq:Re_G_ansatz}). Also, the RPA dispersion relation is
shown, with $1\,-\,V(k)\,\chi^{(0)}(k,\omega(k))=0$ for $k<k_0$ and 
the position of the maximum of $\mbox{Im}\,\chi^{\rm RPA}_{\rm
  ee}(k,\omega)$ for $k>k_0$. 

The shift at $k=0$ shown in these figures corresponds to the values 
for $r_s=1,2,5$ in Fig.~\ref{fig:plasmon_k=0_nu}. In this limit, the 
local field corrections do not contribute. Thus, the different
approximations merge for $k \to 0$.
As in the long-wavelength limit, the deviation of all approximative expressions
from the RPA results are more pronounced with increasing $r_s$. 
As for the Born-Mermin result, it shows a systematic behavior with
a switch from a blue shift to a red shift at a value of $k$ close to
$k_0$.
Similar results have been reported by Thiele et
al. \cite{Thiele08}, where calculations for finite temperature
conditions  at moderate degeneracy are given. We refer for details to that 
paper and take the traditional Born-Mermin results as a reference
point for the inclusion of local field effects. 
Using the extended Born-Mermin
approach together with a static local field correction
shows a drastic change in the
dispersion relation for larger values of $k$. This is expected from 
Fig.~\ref{fig:static_local_fields}, where $G_{ee}(k)$ shows 
considerable deviations from the RPA limit, i.e. $G_{ee}(k)=0$. Also,
the larger local field factor for increasing $r_s$ leads to a more
pronounced change in the dispersion relation, as can be seen by
comparing the three values of $r_s$. However, once we refine the
approximation by allowing for dynamic local field corrections, these 
drastic changes disappear again and a dispersion close to the original 
Born-Mermin curve is found, at least for $r_s=1$ and $r_s=2$. In the
case of $r_s=5$, noticeable differences from both, BM and BM+sLFC remain.
This behavior can be understood by inspection of the frequency
dependence of $\mbox{Re} \,G(k,\omega)$. While $\mbox{Re}\,G(k,0)$ 
increases with $k$, $\mbox{Im}\, G(k,\infty)$ is considerably smaller or
even negative for large $k$, see e.g. \cite{Iwamoto84}. As a
consequence, values at intermediate frequencies are considerable
smaller than the static value $\mbox{Re}\,G(k,0)$. For larger values
of $k$, a zero is even found corresponding to a RPA-like behavior at this
frequency. We illustrate this fact in Fig.~\ref{fig:Re_Im_G_k_omega}, 
showing the 
real and the imaginary part of the dynamic local field correction
as a function of the frequency $\omega$ for $k=k_F$ and $r_s=2$.
Note however, that the dispersion relation is somewhat misleading. 
A frequency scan of the dynamic structure factor reveals a rich
structure, which can not adequately represented by a shift 
and a width only.

Finally, we investigate the plasmon width at $r_s=5$ for the three
different approximations. Again, at $k=0$, the broadening is only due
to the collision frequency. For $k > k_0$, 
the rapid increase of the broadening with $k$ is due to the RPA contribution.
At finite k, local field effects also contribute.  
In particular, the imaginary part of the dynamic local field
correction  adds to the total width of the
plasmon, if one compares the BM + dLFC result to the BM curve.
However, the net effect is not simply the sum of both contributions
due to the involved arithmetics of the Mermin expression. This also
is revealed by the BM + sLFC result, which deviates noticeably from the
BM although no additional imaginary part for $G(k)$ is taken into account.

\section{Conclusions}
\label{sec:conclude}
We devise an interpolation scheme which incorporates electron-ion
collisions as well as electron-electron correlations by an extended 
Mermin approach. We apply this to an interacting electron gas at $T=0$
interacting with an inert ion background. As inputs act  a dynamical 
collision frequency in Born approximation and different models for 
dynamic local field corrections of the interacting OCP electron gas.
Plasmon properties serve as a probe for the relevance of collisions 
and correlations, respectively.

At small wave vectors, we observe a dominance of collisions.
The importance of local field corrections increase with 
increasing Brueckner parameter $r_s$. For $r_s=5$ it is
indispensable to account for local field correlations.
Drastic changes in the plasmon properties which occur by
taking account of static local field corrections disappear 
to some extent when using models for dynamic local field corrections.
The plasmon broadening shows in general a very involved behavior,
collisions and corrections to do not simply add up. 

To apply our model to a realistic situation of Thomson scattering 
in cold matter, we have to refine the input quantities such as the
collision frequency and the model for the OCP local field corrections.
At this point, there is certainly some improvement to be done.

Our approach can also serve as an approach to study the interplay of
impurity scattering and electron-electron correlation in a jellium
model description of metals. Here, connection can be made to
experimental results for the dynamic structure factor measured by 
inelastic x-ray scattering \cite{IXS}.

\begin{acknowledgments}
 
This work was supported by the Deutsche Forschungsgemeinschaft (DFG) within
the Sonderforschungsbereich SFB 652.

\end{acknowledgments}


\begin{thebibliography}{5}
\bibitem{Anmerkung} In difference to \cite{Hong85}, 
                    a factor of $z$ has been included into the
                    definition of $c_2^0$. Furthermore, $-i z$ is used
                    instead of $z$.
\end{thebibliography}
\end{document}